\documentclass[conference]{IEEEtran}
\IEEEoverridecommandlockouts

\usepackage{cite}
\usepackage{amsmath,amssymb,amsfonts}
\usepackage{graphicx}
\usepackage{textcomp}
\usepackage[table,dvipsnames]{xcolor}
\usepackage[hyphens]{url}
\usepackage{makecell}

\usepackage{svg}
\usepackage{bm}
\usepackage{tikz}

\usepackage{algorithm}
\usepackage[noend]{algpseudocode}
\usepackage{afterpage}
\usepackage{enumitem}
\usepackage{multirow}
\usepackage{multicol}
\usepackage{graphicx}
\usepackage{hyperref} 

\usepackage{lipsum}
\usepackage{tabularx}
\usepackage{array}
\usepackage{booktabs}
\usepackage{listings}
\usepackage{framed}
\usepackage{amsthm}
\newtheorem{definition}{Definition}
\usepackage{caption}

\def\BibTeX{{\rm B\kern-.05em{\sc i\kern-.025em b}\kern-.08em
    T\kern-.1667em\lower.7ex\hbox{E}\kern-.125emX}}
\begin{document}

\pdfpagewidth=8.5in
\pdfpageheight=11in

\newcommand{\iscasubmissionnumber}{1962}

\pagenumbering{arabic}

\title{
  A Vertically Integrated Framework for Templatized Chip Design
  }

\author{
\IEEEauthorblockN{Jeongeun Kim and Christopher Torng}
\IEEEauthorblockA{Department of Electrical and Computer Engineering,
  University of Southern California, Los Angeles, CA}
}

\maketitle
\thispagestyle{plain}
\pagestyle{plain}


\begin{abstract}
Developers who primarily engage with software often struggle to incorporate custom hardware into their applications, even though specialized silicon can provide substantial benefits to machine learning and AI, as well as to the application domains that they enable. This work investigates how a chip can be generated from a high-level object-oriented software specification, targeting introductory-level chip design learners with only very light performance requirements, while maintaining mental continuity between the chip layout and the software source program. In our approach, each software object is represented as a corresponding region on the die, producing a one-to-one structural mapping that preserves these familiar abstractions throughout the design flow.
To support this mapping, we employ a modular construction strategy in which vertically composed IP blocks implement the behavioral protocols expressed in software. A direct syntactic translation, however, cannot meet hardware-level efficiency or communication constraints. For this reason, we leverage formal type systems based on sequences that check whether interactions between hardware modules adhere to the communication patterns described in the software model. We further examine hardware interconnect strategies for composing many such modules and develop layout techniques suited to this object-aligned design style. Together, these contributions preserve mental continuity from software to chip design for new learners and enables practical layout generation, ultimately reducing the expertise required for software developers to participate in chip creation.
\end{abstract}

\section{Introduction}
\label{sec-intro}

Chip design demands expertise ranging from logical hardware description and verification, to microarchitectural design metric trade-off analysis, to physical design and implementation~\cite{ajayi-openroad-dac2019, witharana-hw-verification-2022survey}. These requirements block software-centric learners, students, and first-time chip designers from building, or even attempting, a working silicon prototype chip, even when the design intent is clear at the software level. Lowering this barrier while explicitly exposing hardware-level and layout-level artifacts creates both practical and educational value: users design in a familiar programming model and still see how their decisions materialize within a chip design plot.
Prior efforts reduce entry cost but fall short of the goal of this work. For example, Chisel~\cite{bachrach-chisel-dac2012} and other productive hardware description languages raise abstraction levels, bringing Scala-level object-oriented and functional constructs into hardware design, yet designers still write a hardware-embedded DSL and reason about design primarily at the RTL abstraction. High-level synthesis translates software into hardware, but it focuses on function-level kernels and throughput-oriented pipelines, not on preserving object-oriented structure and inter-object protocols all the way down to layout~\cite{ye-hls-2023, cong-hls-fpga-2022}. Literature is relatively sparse on work exploring how to bridge from unmodified object-oriented software programs and connect to manufacturable chip layouts, especially while reflecting the same object-oriented structure, while keeping the mapping formally accountable at every step.

The goal of this work is to advance the frontier of building a road from software-first code to working chips, and to pull upwards a restricted, discretized, but very real controllability of chip design optimizations into the hands of its users.
Our framework, called Marionette, advances a software-first approach that starts with objects. Users write a standard object-oriented program. Objects interact through well-defined protocols. Marionette preserves the object abstraction into hardware and then through to layout without users switching languages or abandoning software modularity. The core mechanism is a session-typed specification of object interactions that serves as a formal contract across the entire vertical stack. Session types are a programming language primitive that formalizes sequences of events between multiple parties into a type~\cite{castagna-session-type-foundations-2009,dardha-sessiontype-2012, dezani-sessiontype-2006, neubauer-sessiontype-implementation-2004, huttel-sessiontype-foundations-2016, honda-sessiontype-multiparty-2008}. We propose using session types to capture the structure and sequencing of communication between object endpoints and to define both the boundary of each object and the communication protocol between objects. Then Marionette couples these specifications with a vertical IP stack (software, RTL, and layout) that maintains the same decomposition at each level: each software object corresponds to an RTL IP and to a layout-level IP template.

Our target users control chip design exclusively from the software level. However, this leads to several two key early challenges. First, achieving the same behavior as in software at the RTL and layout levels is often undesirable, because naive replication of the software communication may be unrealistic or low performance in hardware. Examples include what it means to pass a string as a first-class message or how to handle a call that transfers a large image (e.g., 4 GB) across object boundaries. Hardware demands width-bounded, latency-aware channels, explicit buffering, and topology-constrained routing. Second, once the communication is hardened to match the requirements of hardware, the resulting hardware designs drift away from the original software objects in both boundary interface specification and behavior, which calls into question whether the system of communicating software objects is equivalent to the new system of differently communicating hardware objects. Marionette addresses these challenges by providing strong, formal statements of: (1)~object internal and external specifications (i.e., what happens inside and what happens outside), and (2)~the protocol specification that must hold between objects at both software and hardware layers. Then these properties are enforced and propagated through from the RTL implementation to the physical interconnect to ensure conforming to the same protocol.

Marionette compiles session-typed object interactions into RTL components and channels that implement the identical protocol semantics. As a result, the designer programs modular objects in software and receives a modular RTL netlist and a layout that mirrors the software structure. This vertical mirroring is deliberate for education: learners observe how object boundaries map to module boundaries and how those in turn map to floorplanning and interconnect.

Productive hardware description approaches such as Chisel lower the barrier to RTL authoring and automates idiomatic hardware patterns; Marionette lowers the barrier further by accepting a subset of natural software programming paradigms and preserving object-level structure and protocols through to layout. Recall that HLS synthesizes computations but is still a hardware-level description, and it does not aim to preserve object boundaries and session-typed interactions as first-class entities across RTL and floorplanning. Marionette’s objective is therefore unique: protocol-first, object-preserving chip realization. Our contributions are as follows:
\begin{itemize}
    \item Session-typed protocol verification across the stack: We present a framework that takes the user's intended object-level communication protocol and verifies its realizability and conformance at the hardware level. Session types define object boundaries and legal interaction traces; Marionette propagates these contracts to RTL channels and layout interconnect and rejects implementations that violate the protocol.
    \item Object-centric exploration of hardware interconnects: We study interconnect organizations for connecting multiple object-derived IP modules so that object-oriented structure scales to systems of interacting blocks. The framework selects and configures interconnect (point-to-point, switched, or hierarchical) to implement the same sessions with hardware-appropriate flow control and backpressure, without altering the object graph.
    \item Layout template library: We study how to produce different physical layout templates that implement the same function in different performance levels (exposing performance optimization to software-first chip designers), and different areas and aspect ratios (exposing silicon utilization and floorplanning).
\end{itemize}

The Marionette framework accepts object-oriented software descriptions, emits protocol-checked RTL, and produces manufacturable chip layouts built from a vertical IP library. 
By elevating protocols to first-class specifications and by aligning software objects with hardware and layout entities, Marionette delivers a formally checkable mapping from software intent to silicon structure.

We hope that our work establishes a foundation that enables software-oriented designers to pursue an end-to-end path to chip prototyping without sacrificing the visibility of familiar software abstractions in their RTL and layout. In doing so, we choose to accept significant sacrifices to design metrics in chip design (performance, power, area), while still outperforming accessible alternatives to our audience (e.g., Arduino-style microcontrollers). We expose chip design optimization in a heavily discretized and templated fashion to smoothen the entry as users delve into custom silicon. We aim to provide educators with a platform in which abstractions and physical realizations are structurally aligned, allowing students and learners to trace design intent consistently across all layers.



\begin{figure}[t]

  \centering
  \includegraphics[width=1\linewidth]
  {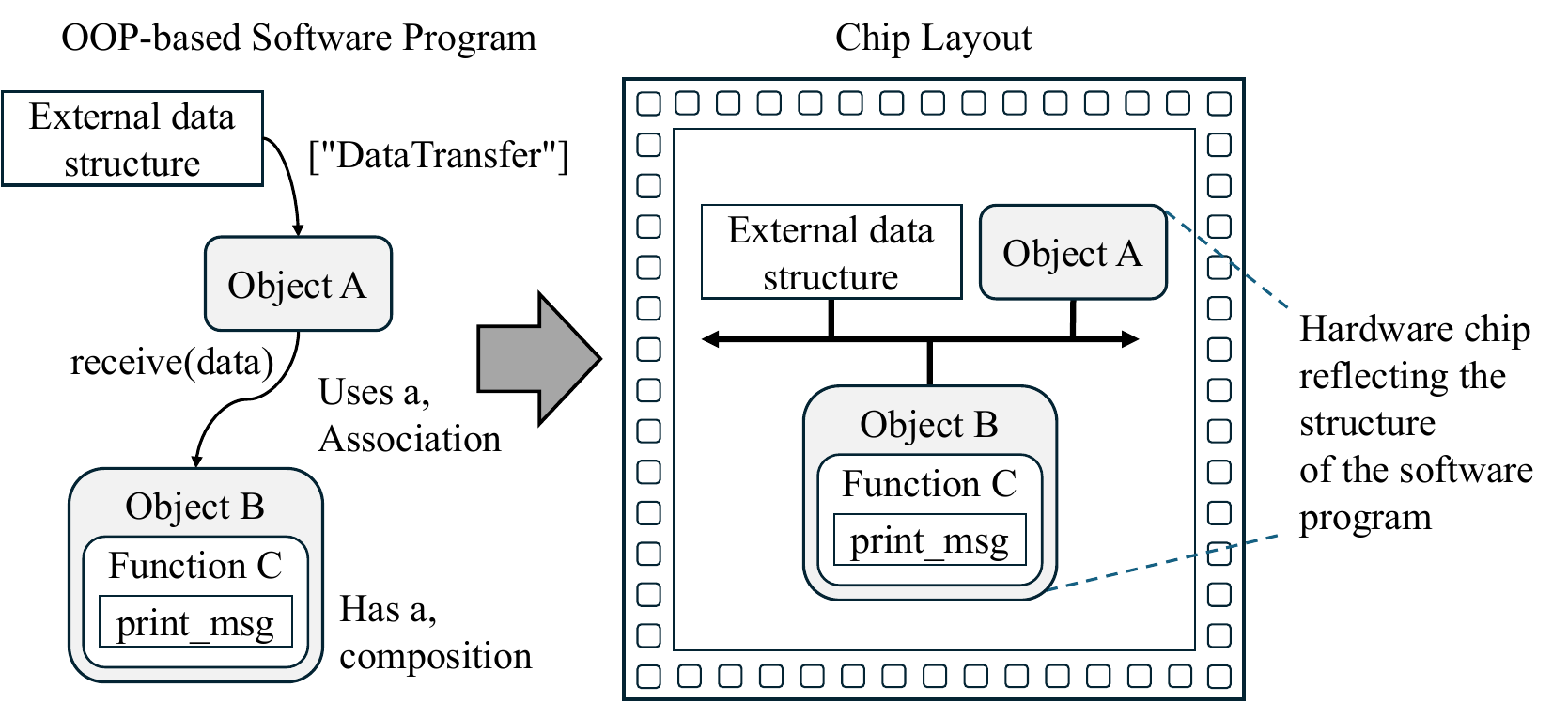}
  \caption{System architecture mapping: the object-oriented structure of the software design is directly reflected in the chip layout.}

  \label{fig-system-goal}
  \vspace{-0.0in}

\end{figure}
\begin{table}[t]
\small 
\centering
\caption{Mapping of Object-Oriented Software Concepts to Hardware}
\label{tbl-sw_hw_mapping}
\begin{tabularx}{\columnwidth}{l l X}

\toprule
\multicolumn{1}{c}{\textbf{OOP Concept}} & 
\multicolumn{1}{c}{\textbf{HW Concept}} \\
\midrule
Function / Method         & Operation unit, FSM action, ALU \\
Class                     & Hardware module, IP block \\
Object (Instance)         & Module instance \\
Data Structure            & Memory structure (registers, SRAM) \\
Inheritance               & Parameterization, module reuse \\
Encapsulation             & Module boundary, internal wires \\
Composition               & Hierarchical module composition \\
\bottomrule
\end{tabularx}
\end{table}


\begin{figure}[t]

  \centering
  \includegraphics[width=0.9\linewidth]{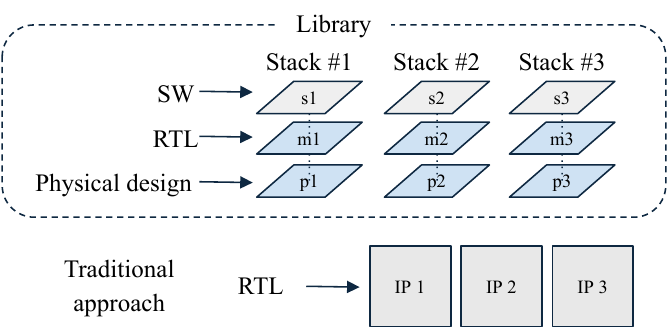}
  \caption{A vertically stacked IP library, integrating software, RTL, and physical design as a single unit, differs from traditional horizontally layered IPs.}

  \label{fig-vertical-ip}
  \vspace{-0.0in}

\end{figure}

\section{System Goals}
\label{sec-system-goals}

The primary goal of the proposed Marionette framework is to enable software developers, without specialized hardware expertise, to design modular chip prototypes directly using familiar programming paradigms. Figure~\ref{fig-system-goal} shows how Marionette takes a system of software objects (left) and returns a system of hardware objects in a manufacturable layout (right).
Table~\ref{tbl-sw_hw_mapping} shows how object-oriented software programming terminology can correspond to hardware contexts.

\textbf{Aspects of Vision Shape our Research Questions of Interest.} There are fundamental questions regarding the goals of a software-first framework that would impact the expected studies for this work. One valid vision is to enable software-first designers to build \textit{fast and efficient hardware}, which would direct our study towards questions on how to convert sequential bottlenecks from software abstractions into the concurrency properties inherent to hardware. While interesting, our mission is instead ambitious along another dimension: pulling upwards a restricted, discretized, but very real \textit{mindset of controllable chip design optimizations} into the hands of software-first users. This means we seek to build a chip design optimization space, even at the cost of placing it at a lower absolute range of design quality metrics. \textbf{Impact:} We focus our research on aspects of this challenge that cannot be solved with greater near-term engineering effort. Specifically, we view systems as composed of objects (i.e., hardware IP blocks) and inter-object communication (i.e., interconnect and message sequences). Producing hardware derived from high-level descriptions is a much more well-studied research frontier in the high-level synthesis space. We see inter-object communication as the primary technical research challenge that cannot be sidestepped by engineering, and we choose to address this aspect in this work.
As a result, the goal and contribution of Marionette is not to develop new hardware-generation techniques at the intra-function level. Instead, our focus is on the fundamental problem of establishing a sound and analyzable correspondence between a software system of interconnected objects and its hardware realization. This leads to the following objectives.

\textbf{Software-Driven Vertical IP Stack and Compositional Formal Equivalence.}
Marionette enables hardware design directly from software by employing vertical IP stacks that bind together a software object, its corresponding RTL description, and its associated physical layout blocks (Figure~\ref{fig-vertical-ip}). In contrast to conventional horizontal IP integration, Marionette structures IPs vertically, allowing software-level communication structures to propagate deterministically to RTL and layout.
We expect that these IPs will eventually be generated automatically through high-level synthesis techniques, but we assume for our research scope that IPs are covered near-term by greater engineering effort (i.e., pre-built IP library blocks).
By integrating independently verified vertical IP stacks without requiring additional top-level transformations, this methodology guarantees functional correctness and ensures DRC-/LVS-clean layouts by construction: intra-block manufacturability is secured through IP design, while inter-block manufacturability is enforced by the framework’s interconnection architecture.
The central technical challenge is maintaining user-defined communication semantics coherently across the software, RTL, and layout layers. Marionette achieves this by ensuring that composition in software induces an equivalent composition in hardware, enabling developers to design entirely at the software level while relying on the framework to materialize correct-by-construction RTL and layout structures. We will formalize and guarantee behavioral correspondence between composed software modules and composed hardware modules.
The example vertical IP stacks we build to demonstrate this workflow are not highlighted contributions of this work.

\textbf{Abstraction with Performance Awareness.}
Marionette preserves the object-oriented hierarchy of user programs through the RTL and layout levels (Figure~\ref{fig-system-goal}), enabling developers to reason about hardware organization directly in terms of their original software structure. This stands in contrast to high-level synthesis (HLS), which abstracts at the function level within a host language and typically obscures structural relationships.
A key challenge is surfacing hardware trade-offs—such as throughput, latency, and area—without requiring specialized EDA knowledge. Marionette addresses this by providing template libraries for each IP module, offering multiple hardware realizations with distinct performance and area characteristics. Developers can thus explore the design space entirely at the software abstraction layer, while the framework abstracts away RTL and layout complexity.

\textbf{Software-Level Chip Prototyping.} Ultimately, Marionette allows developers to obtain modular chip prototypes with tunable performance and area by writing conventional software code. Optimization decisions, traditionally confined to hardware experts, become accessible at the software abstraction.


\section{Marionette Framework}
\label{sec-marionette}


\begin{figure*}[t]

  \centering
  \includegraphics[width=1\linewidth]{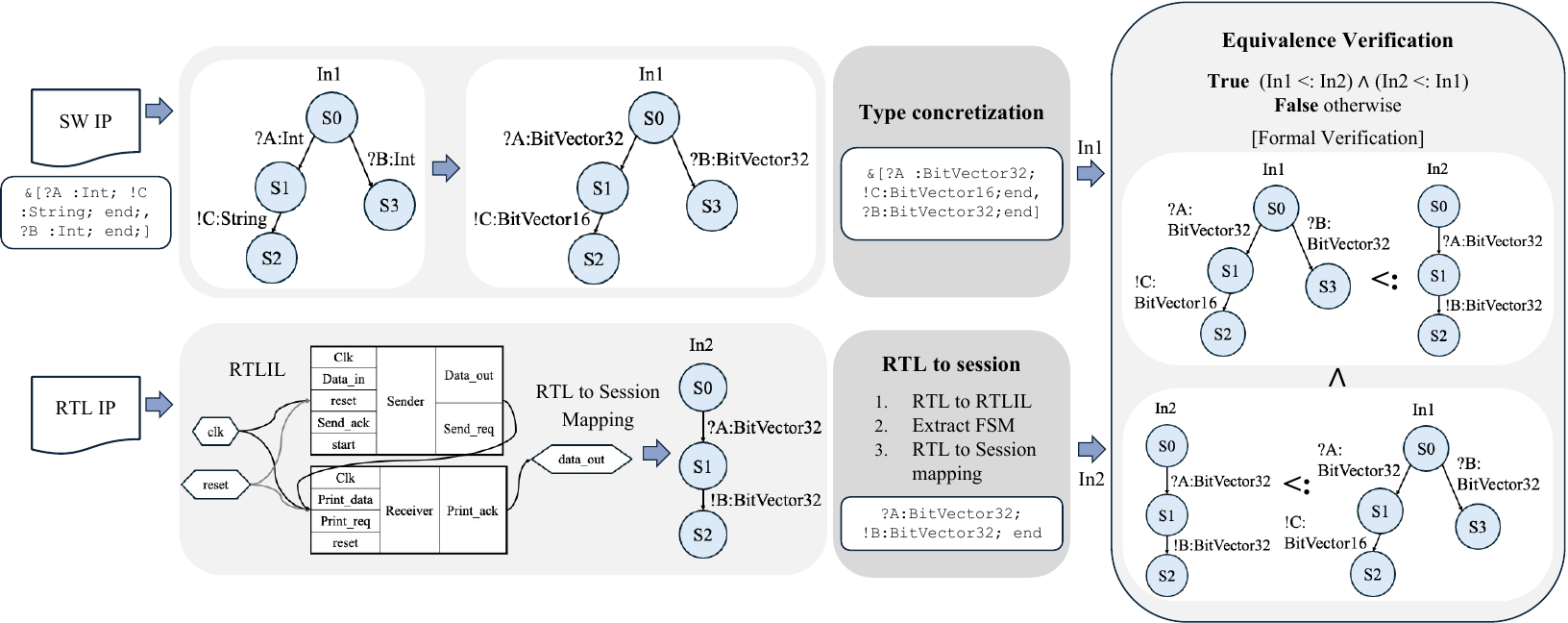}
  \caption{High-level equivalence verification flow: starting from a vertical IP, only connections between modules that pass inter-layer communication protocol equivalence checks are realized in the chip layout.}

  \label{fig-high-level-flow}
  \vspace{-0.0in}

\end{figure*}

The Marionette framework can be decomposed into its contributions at each pair of layers. At the software-to-RTL layer, we present a sequence-based formal equivalence checking approach built upon session types to prove that the system of software objects communicates in the same way as the system of hardware objects (Section~\ref{sec-equivalence}). This is the primary contribution of our work. At the microarchitecture and RTL layer, we contribute a study on how different interconnection network architectures can trade off maintaining software-level communication sequences in exchange for optimizing hardware metrics (Section~\ref{sec-network}). Finally, our contribution at the RTL-to-layout layer is a physical template library that enables a software-first chip designer to optimize for performance, area, and silicon utilization in a friendly graphical manner (i.e., with a GUI), thereby opening the black box and exposing simple optimizations to entry-level, software-first chip designers (Section~\ref{sec-template-library}).

\label{sec-methodology}

\subsection{Equivalence Verification Framework Overview}
\label{sec-equivalence}

Marionette leverages session types~\cite{gay-sessiontype-linear-2010, kouzapas-session-type-2024lmcs, das-work-session-type-2018lcs, marques-session-type-verification-2013arxiv} to formally specify software-level interactions, enabling static verification of protocol correctness. By enforcing communication compatibility, the framework guarantees that only modules with mutually consistent behaviors are composed, and that these guarantees are preserved through RTL synthesis and layout generation. With session types as a formal foundation, Marionette provides a unified mechanism to capture both sequencing and data constraints, maintaining communication behavior through the translation from software to hardware.

Figure~\ref{fig-high-level-flow} summarizes how Marionette verifies cross-layer equivalence. The framework represents software-level interactions as labeled transition system (LTS) graphs~\cite{bacchiani-session-subtyping-tool-2021} annotated with payload bit-widths. On the hardware side, RTL descriptions of the corresponding IPs are compiled into dataflow graphs. Communication points extracted from these RTL graphs are then lifted into LTS form using session types, yielding a hardware-level representation of communication behavior.
With both layers expressed as comparable LTS graphs, Marionette performs equivalence checking to determine whether a hardware communication pattern matches the software-specified protocol. If a valid RTL-level connection is found, the framework proceeds to layout generation, producing a modular physical design. If no equivalent communication behavior exists in hardware, Marionette flags a protocol violation, indicating that the intended software-level interaction is unrealizable under the given IP modules.

\label{sec-valid-program-input}

\begin{figure}[t]

  \centering
  \includegraphics[width=1\linewidth]{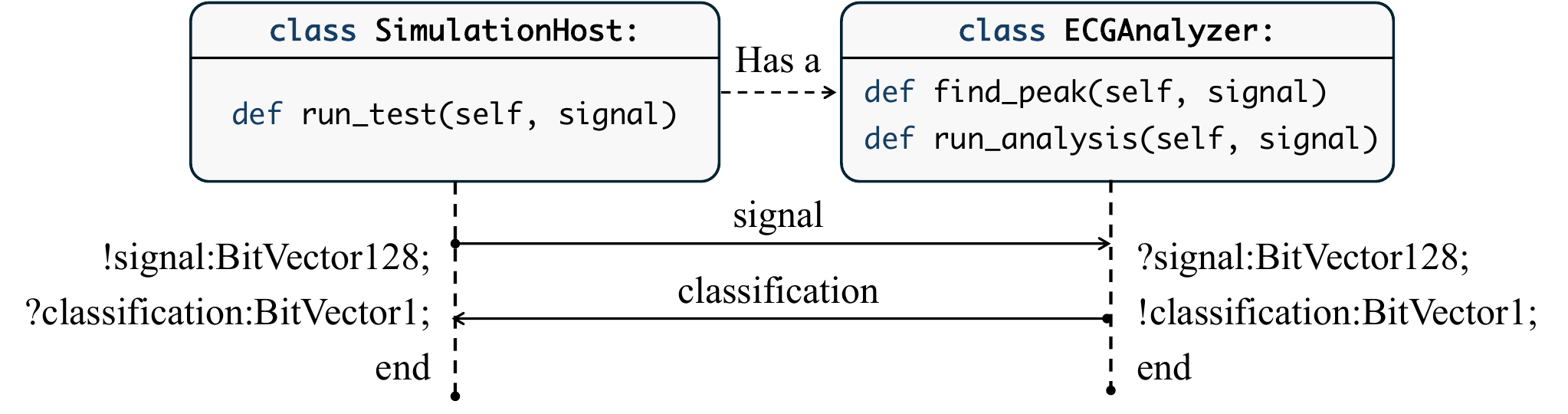}

  \caption{Example of user software input and the corresponding session type sequence.}

  \label{fig-input-example}
  \vspace{-0.0in}

\end{figure}

\subsection{Limitations of the Session Type Approach}

Session types in Marionette provide a formal guarantee of equivalence between software-level object interactions and their hardware implementation. To enforce this guarantee, Marionette imposes a structured input specification: (i) connections between software modules must adhere to session type protocols, (ii) the corresponding RTL IPs must provide FSM semantics, and (iii) each FSM state is restricted to a single session-type action (send, receive, choose, or offer) to enable protocol verification.
Type systems fundamentally capture what behaviors are and are not allowed, restricting which programs are and are not expressible. For Marionette, our use of a type system is to construct an environment in which formal equivalence checking is possible, and we accept the resulting reality of defining a restricted expressible program space, in the same way that SystemVerilog has a restricted synthesizable subset. Our constraints are as follows:

First, dynamic Python features, such as variable-length lists and append operations, are not directly supported. All messages exchanged between objects must have statically defined lengths and types. For example, Python lists must be converted to fixed-size arrays, and all data types such as string, integer must be mapped to fixed-size representations suitable for hardware implementation.

Second, sequential execution constraints are strictly enforced. Only one input set may be transmitted at a time, and concurrent or multi-threaded object-to-object communication is disallowed in this work. Any deviation from the expected message order or type results in session termination, ensuring deterministic and synthesizable hardware behavior.

\textbf{Example of User Program Inputs}

As shown is Figure~\ref{fig-input-example}, the following Python example with session type is drawn from a project conducted by first- and second-year undergraduate students, who represent typical Marionette target users, and analyzes ECG signals to diagnose conditions such as tachycardia.

\label{code-main}



\lstdefinestyle{mypython}{
    language=Python,
    backgroundcolor=\color{gray!2}, 
    basicstyle=\ttfamily\scriptsize, 
    keywordstyle=\color{blue}\bfseries,
    commentstyle=\color{green!50!black}\itshape,
    stringstyle=\color{red!80!black},
    numbers=left,
    numberstyle=\tiny\color{gray},
    stepnumber=1,
    numbersep=6pt,
    frame=none,
    breaklines=true,
    breakatwhitespace=true,
    showstringspaces=false,
    xleftmargin=2em
}

\begin{figure}[t]
    \centering
    \begin{lstlisting}[style=mypython]
if __name__ == "__main__":
    # Instance  
    analyzer = ECGAnalyzer()
    host = SimulationHost(analyzer)

    # Valid example
    signal = [1, 6, 8, 5] # 4-element array

    # Valid method call
    print(host.run_test(signal))
    \end{lstlisting}
    \caption{Valid software input example: program entry point with fixed-size input and valid method calls.}
    \label{fig-code-main}
\end{figure}

\label{code-simhost}




\lstdefinestyle{mypython}{
    language=Python,
    backgroundcolor=\color{gray!2}, 
    basicstyle=\ttfamily\scriptsize, 
    keywordstyle=\color{blue}\bfseries,
    commentstyle=\color{green!50!black}\itshape,
    stringstyle=\color{red!80!black},
    numbers=left,
    numberstyle=\tiny\color{gray},
    stepnumber=1,
    numbersep=6pt,
    frame=none,
    breaklines=true,
    breakatwhitespace=true,
    showstringspaces=false,
    xleftmargin=2em
}

\begin{figure}[t]
    \centering
    \begin{lstlisting}[style=mypython]
class SimulationHost:
    def __init__(self, analyzer):
        self.analyzer = analyzer

    def run_test(self, signal):
        classification = self.analyzer.run_analysis(signal)
        classification
    \end{lstlisting}
    \caption{\texttt{SimulationHost} calls \texttt{run\_analysis} of \texttt{ECGAnalyzer}, passing a signal and obtaining the classification result}
    \label{fig-code-simhost}
\end{figure}

\label{code-ecg}




\lstdefinestyle{mypython}{
    language=Python,
    backgroundcolor=\color{gray!2}, 
    basicstyle=\ttfamily\scriptsize, 
    keywordstyle=\color{blue}\bfseries,
    commentstyle=\color{green!50!black}\itshape,
    stringstyle=\color{red!80!black},
    numbers=left,
    numberstyle=\tiny\color{gray},
    stepnumber=1,
    numbersep=6pt,
    frame=none,
    breaklines=true,
    breakatwhitespace=true,
    showstringspaces=false,
    xleftmargin=2em
}

\begin{figure}[t]
    \centering
    \begin{lstlisting}[style=mypython]
class ECGAnalyzer:
    def find_peak(self, signal):
        return max(signal)

    def run_analysis(self, signal):
        peak = self.find_peak(signal)

        # Dummy HR rule based on peak
        heart_rate = 72 if peak < 1.0 else 120
        classification = "Normal" if heart_rate < 100 else "Tachycardia"
        return classification
    \end{lstlisting}
    \caption{Simplified \texttt{ECGAnalyzer} processing a signal to produce a classification result}
    \label{fig-code-ecg}
\end{figure}

\lstdefinestyle{mypython}{
    language=Python,
    backgroundcolor=\color{gray!5}, 
    basicstyle=\ttfamily\scriptsize, 
    keywordstyle=\color{blue}\bfseries,
    commentstyle=\color{green!50!black}\itshape,
    stringstyle=\color{red!80!black},
    numbers=left,
    numberstyle=\tiny\color{gray},
    stepnumber=1,
    numbersep=6pt,
    frame=none,
    breaklines=true,
    breakatwhitespace=true,
    showstringspaces=false,
    xleftmargin=2em
}

\begin{figure}[h]
    \centering
    \begin{lstlisting}[style=mypython]
def call_host(host, signal):
    print(host.run_test(signal))

if __name__ == "__main__":
    # Instance
    analyzer = ECGAnalyzer()
    host = SimulationHost(analyzer)

    # Variable-length, dynamic append
    signal_invalid = [1, 6]   # not fixed-length
    signal_invalid.append(9)  # dynamic append

    # Concurrent method calls
    t1 = Thread(target=call_host, args=(host, signal_invalid))
    t2 = Thread(target=call_host, args=(host, signal_invalid))
    t1.start()
    t2.start()
    t1.join()
    t2.join()
    \end{lstlisting}
    \caption{Invalid software input example: Marionette does not support dynamic append operations or concurrent method calls}
    \label{fig-code-main-invalid}
\end{figure}

Figures~\ref{fig-code-main}, \ref{fig-code-ecg}, and \ref{fig-code-simhost} illustrate a user program written in an object-oriented style. In this program, \texttt{SimulationHost} and \texttt{ECGAnalyzer} communicate by sending ECG signal data and receiving analysis results. Marionette assumes hardware modularization at the object level, mapping object interactions to sequential FSMs.

\paragraph{Valid Input}  
Figure~\ref{fig-code-main} line~7 shows a fixed-size array with synthesizable types. The sequence of object interactions is strictly sequential, enabling deterministic mapping to FSM states. For instance, \texttt{host.run\_test} calls \texttt{ECGAnalyzer.run\_analysis}, which internally invokes \texttt{find\_peak} and returns a classification. This deterministic order is directly representable as a session type and mappable to FSM states.

\paragraph{Invalid Input}  
Figure~\ref{fig-code-main-invalid} line~11 shows a dynamic list with append operations that violate session type constraints. As shown in line~13, concurrent interactions where multiple objects are accessed simultaneously also violate session type constraints. Hardware FSMs cannot unambiguously represent simultaneous method calls without additional synchronization, making such programs unsupported in Marionette.


\begin{table}[t]
\centering
\small
\caption{Correspondence between RTL Patterns and Session Type Actions}
\label{tbl-rtl-to-session}
\begin{tabular}{c c} 
\toprule
\textbf{RTL Pattern} & \textbf{Session Type Action} \\
\midrule
\texttt{output handshake = 1} & \texttt{!} send \\
\texttt{if (input handshake)} & \texttt{?} receive \\
\texttt{internal branch} & $\oplus$ choose \\
\texttt{input-data-controlled branch} & $\&$ offer \\
\bottomrule
\end{tabular}
\end{table}

\subsection{High-Level Abstraction Mapping}

This section describes how software-level communication behaviors are systematically compared with their hardware-level RTL counterparts. Table~\ref{tbl-rtl-to-session} illustrates an example mapping used to achieve this correspondence. A primary challenge arises from the fundamentally different abstraction levels, which precludes direct comparison.

As illustrated in Figure~\ref{fig-high-level-flow}, obtaining analogous sequences from RTL requires translating the RTL into a dataflow graph (DFG) through the RTL Intermediate Language (RTLIL). From the DFG, a FSM is extracted to capture control flow, signal ordering, and send/receive sequences. All feasible FSM execution paths are enumerated and converted into session-type sequences, forming candidate protocols for cross-layer equivalence verification. Each sequence corresponds to primitives \texttt{send}, \texttt{recv}, \texttt{choose}, and \texttt{offer}: output-valid states map to \texttt{send}, input-valid states to \texttt{recv}, input-driven branching to \texttt{offer}, and internal decisions to \texttt{choose}.

An \textit{LTS} is defined as a tuple:
\[
\mathcal{L} = (S, A, T, s_0)
\]
where $S$ is a finite set of states, $A$ a finite set of labeled actions, $T \subseteq S \times A \times S$ the transition relation, and $s_0 \in S$ the initial state. Each transition $(s_i, a, s_j) \in T$ represents message transmission, reception, or synchronization, capturing the communication context of the protocol. Software-level session types are directly mapped to LTSs, while RTL-derived session types are extracted through a dedicated procedure to ensure semantic fidelity.
To bridge the abstraction gap, RTL-derived LTSs are pruned to retain only communication-relevant states, transitions, and payload metadata, while software-derived LTSs are concretized by annotating abstract types (\texttt{int}, \texttt{bool}, \texttt{string}) with explicit maximum bit-widths to match RTL granularity. Formally, each transition label is represented as a tuple:
\[
\ell = (a, m, \tau, w),
\]
where \(a \in \{\text{! (send)}, \text{? (receive)}, \oplus \text{ (choose)}, \& \text{ (offer)}\}\) denotes the session-type action, \(m\) represents the message identifier, \(\tau\) is the data type of the message, and \(w\) indicates its bit-width. This normalized representation aligns software- and hardware-derived session-type graphs, providing a rigorous basis for filtering candidate RTL connections and enabling systematic verification of cross-layer behavioral equivalence.


\begin{figure}[t]

  \centering
  \includegraphics[width=1\linewidth]{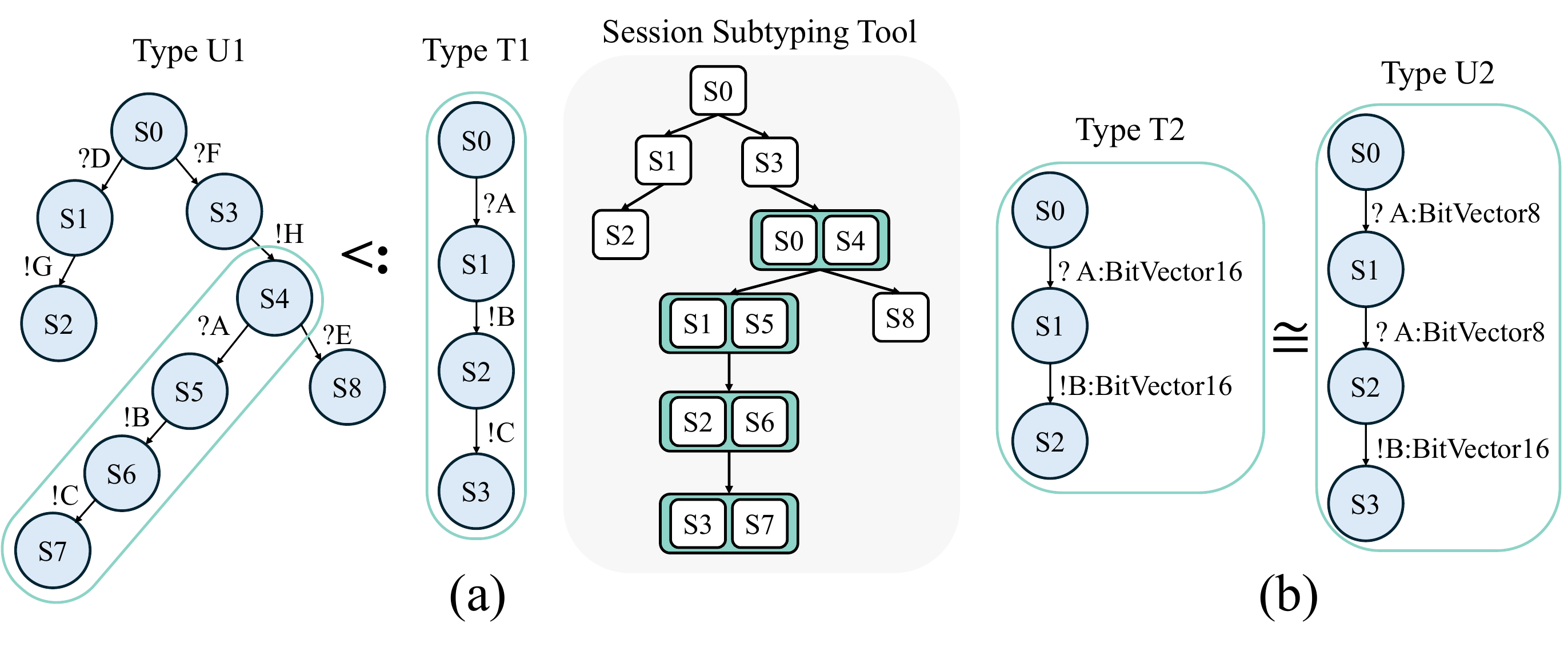}
  \caption{(a) Example of subtyping and the corresponding verification tool.
(b) The necessity of sequence merging arises in behavioral equivalence verification of session-type sequences. Owing to intrinsic differences between software and hardware communication, operations such as \texttt{recv(16-bit A)} in software must be regarded as behaviorally equivalent to two consecutive \texttt{recv(8-bit A)} operations in hardware.}

  \label{fig-subtyping}
  \vspace{-0.0in}

\end{figure}

\subsection{Equivalence Checking via Session Type Subtyping}

Figure~\ref{fig-subtyping}(a) illustrates that, once software- and hardware-derived communication behaviors are normalized as LTS representations, equivalence can be evaluated at a unified abstraction level. We assess whether the two session-type specifications preserve identical communication behavior by leveraging the formal structure of session types. In particular, session-type subtyping is employed to verify that the hardware-derived LTS implements the communication semantics defined at the software level, enabling behavioral verification.
Subtyping provides a formal mechanism to compare two communication protocols by determining whether one protocol can safely replace another without altering the overall system behavior~\cite{gay-subtyping-2005, chen-subtyping-preciseness-2014, ghilezan-subtyping-precise-2023}. Formally, subtyping is denoted as $T <: U$, meaning that type $T$ can be used in any context where type $U$ is expected.
Consider two protocols represented as LTS, shown in Figure~\ref{fig-subtyping}(a). The first protocol, $U_1$, can be defined, after extracting the core branching structure, as:
\vspace{-1mm}
\[
\begin{gathered}
\mathcal{L}_{U_1} = (S_{U_1}, A_{U_1}, T_{U_1}, s_4),\\
S_{U_1} = \{s_4, s_5, s_6, s_7, s_8\},\\
A_{U_1} = \{recv(A), send(B), send(C), recv(E)\}.
\end{gathered}
\]

Its transitions introduce an external choice: from the initial state, the protocol may either receive \(\texttt{A}\) (following the same sequence as $T_1$) or receive \(\texttt{E}\) and terminate.
The second protocol is defined as:
\vspace{-1mm}
\[
\begin{gathered}
\mathcal{L}_{T_1} = (S_{T_1}, A_{T_1}, T_{T_1}, s_0),\\
S_{T_1} = \{s_0, s_1, s_2, s_3\},\\
A_{T_1} = \{recv(A), send(B), send(C)\}.
\end{gathered}
\]

Its transitions form a linear sequence: the protocol first receives \(\texttt{A}\), then sends \(\texttt{B}\), and finally sends \(\texttt{C}\).
In this scenario, $U_1$ is a subtype of $T_1$ (i.e., $U_1 <: T_1$). This relationship adheres to the principle of safe substitution: a context designed for $T_1$ will only send message \(\texttt{A}\), which $U_1$ can handle, ensuring safe interaction. Conversely, a context designed for $U_1$ might send \(\texttt{E}\); substituting $T_1$ in this case would result in a protocol unable to process the message, leading to failure. This example illustrates the subtyping rule for external choice, grounded in the contravariance of inputs: a subtype must be able to handle a superset of the inputs expected by its supertype. Accordingly, two session types are considered behaviorally equivalent if they are mutual subtypes.

\begin{definition}[Session Type Equivalence]
Given two session types $T$ and $U$, they are behaviorally equivalent, denoted $T \approx U$, if and only if both $T <: U$ and $U <: T$.
\end{definition}

By this definition, mutual subtyping guarantees that either protocol can safely substitute for the other in any valid communication context.


\subsection{Data-aware Equivalence Verification}


Consider the following simple session type:
\[
send(image) \texttt{$\rightarrow$} recv(ack)
\]
which can be represented as an LTS with:
\begin{itemize}
    \item States: $S = \{s_0, s_1, s_2\}$
    \item Actions: $A = \{send(image), recv(ack)\}$
    \item Transitions: $(s_0, send(image), s_1)$, $(s_1, recv(ack), s_2)$
\end{itemize}

In a corresponding RTL implementation, the \texttt{send(image)} action may be decomposed into multiple serialized transfers over several clock cycles, whereas the software-level abstraction assumes atomicity, transferring the entire image in a single step. At first glance, these behaviors appear different; however, they can be proven equivalent under proper analysis. Establishing this equivalence is essential for ensuring behavioral consistency across abstraction layers.
As illustrated in Figures~\ref{fig-high-level-flow} and ~\ref{fig-subtyping}(b), unlike traditional session-type approaches that abstract or oversimplify data, our model preserves annotations not only for session-type actions but also for their associated data types and messages, thereby enabling precise equivalence verification.
To reconcile the layers, we adopt the following procedure: for each pair of transitions in the software- and hardware-derived graphs, if both actions and data payloads match (Figure~\ref{fig-subtyping}(b)), the sequences are merged and registered as candidate protocols for comparison against the software-level session sequence. This folding mechanism allows multiple hardware-level transitions to be represented as a single high-level action, mitigating behavioral discrepancies arising from structural differences between hardware and software.



\begin{figure}[t]

  \centering
  \includegraphics[width=1\linewidth]{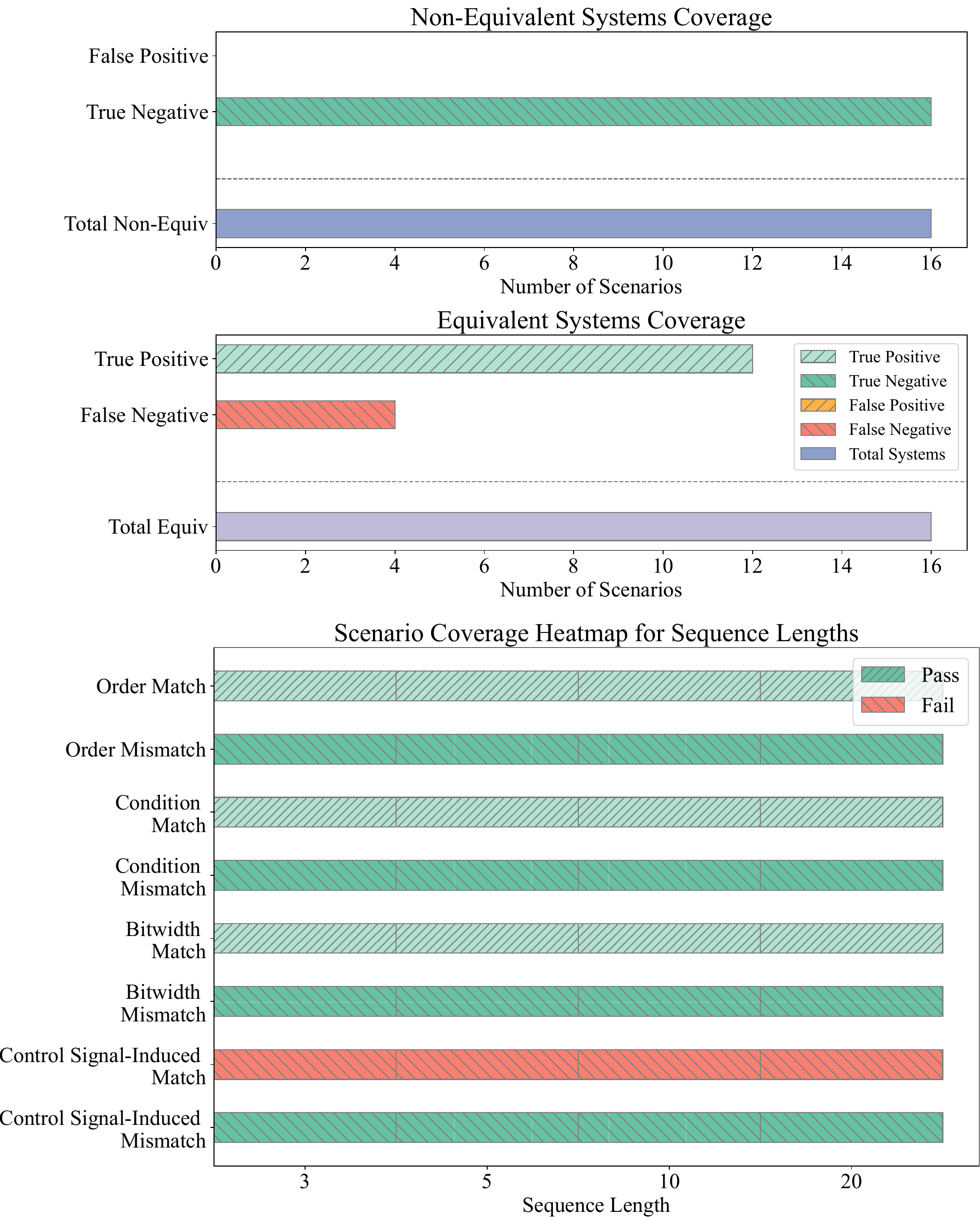}

  \caption{System coverage across \textit{non-eq} and \textit{eq} cases. The figure shows the distribution of false positives (FP), true negatives (TN), true positives (TP), and false negatives (FN), along with a scenario coverage heatmap illustrating how results scale with session sequence complexity.}

  \label{fig-system-coverage}
  \vspace{-0.0in}

\end{figure}

\begin{figure}[t]

  \centering
  \includegraphics[width=1\linewidth]{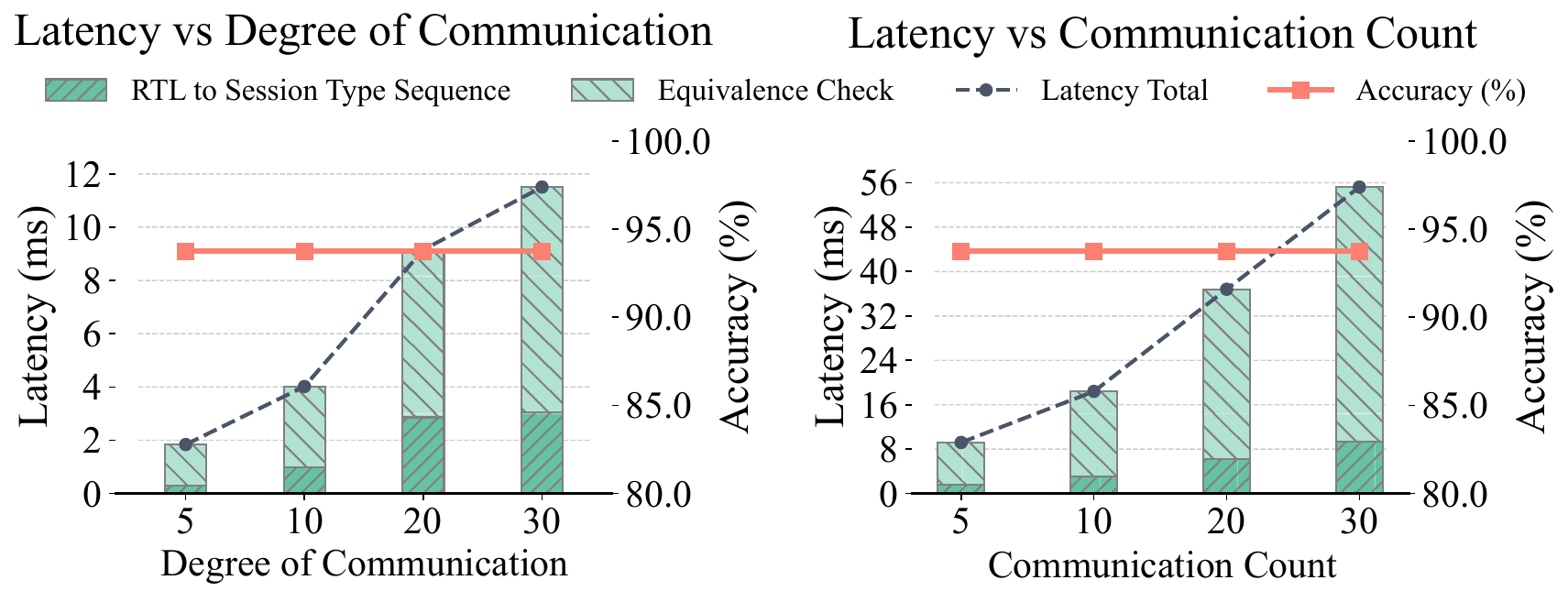}
  \caption{(Left) Equivalence check latency versus the degree of communication, with accuracy maintained. (Right) Equivalence check latency versus total communication count, assuming each communication has a degree of 5, with accuracy maintained.}

  \label{fig-accuracy}
  \vspace{-0.1in}

\end{figure}

\subsection{Scalable Implementation of Behavioral Equivalence Verification}

In this section, we evaluate both the correctness and scalability of our implementation. Within the OpenROAD flow~\cite{yosys-web}, RTL code was synthesized using Yosys into RTLIL and DFG representations, from which FSM sequences were subsequently extracted. For practical verification, we utilized session type subtyping tools~\cite{bacchiani-session-subtyping-tool-2021, bravetti-subtyping-2021lmcs}.

Figure~\ref{fig-system-coverage} summarizes checker outcomes as true positives (TP), true negatives (TN), false positives (FP), and false negatives (FN). No false positives were observed, ensuring that the framework never accepts an incorrect hardware implementation. A limited number of false negatives were observed, primarily due to control signals such as \texttt{start} and \texttt{done} in the hardware IP. These signals, which have no software-level counterparts, are currently treated as data, resulting in spurious transitions (e.g., \texttt{recv(start)}, \texttt{send(done)}). Although these issues impose some limitations on the implementation of hardware IP, they can be addressed in future extensions.

To systematically evaluate coverage, we designed test scenarios that vary session type sequence ordering, message identity, branching, and bit-width mismatches, as well as scenarios that induce control signals, as illustrated in Figure~\ref{fig-system-coverage}. Results show that verification accuracy remains stable as sequence length scales from 3 to 20. When mismatches are intentionally introduced, the checker consistently produces TN outcomes. In contrast, when a hardware session includes additional control signals, the framework may mistakenly classify them as false negatives. While this behavior highlights potential areas for refinement, it does not risk faulty hardware acceptance.

In summary, we have produced a formal equivalence checking tool that corresponds a system of software objects and a system of hardware objects, that:
\vspace{-1mm}
\begin{itemize}
    \item Classifies a subset of equivalent software and hardware systems correctly (i.e., true positive)
    \item Never accepts non-equivalent systems as a match (i.e., false positive), which would allow for incorrect RTL to be taped out as a chip
    \item Classifies a subset of equivalent software and hardware systems as non-equivalent (i.e., false negative), which is only a performance concern
    \item Always rejects non-equivalent systems (i.e., true negatives)
\end{itemize}

\subsubsection*{Constraints Impacting Scalability and Performance}
As shown in Figure~\ref{fig-accuracy} (left), increasing the number of back-and-forth interactions between objects results in longer FSM sequences and extended session-type sequences, causing RTL-to-sequence translation time to scale with sequence length. Figure~\ref{fig-accuracy} (right) shows that a higher total number of communications increases the number of sessions to be checked, raising computational and storage requirements, although verification accuracy remains unaffected.

Overall, the session-type approach enforces strict sequential ordering and explicit type constraints. While these restrictions are essential for correct hardware mapping, they inherently limit concurrency and present scalability challenges. Despite these limitations, the method ensures behavioral equivalence between software and hardware implementations.


\begin{figure}[t]

  \centering
  \includegraphics[width=1\linewidth]{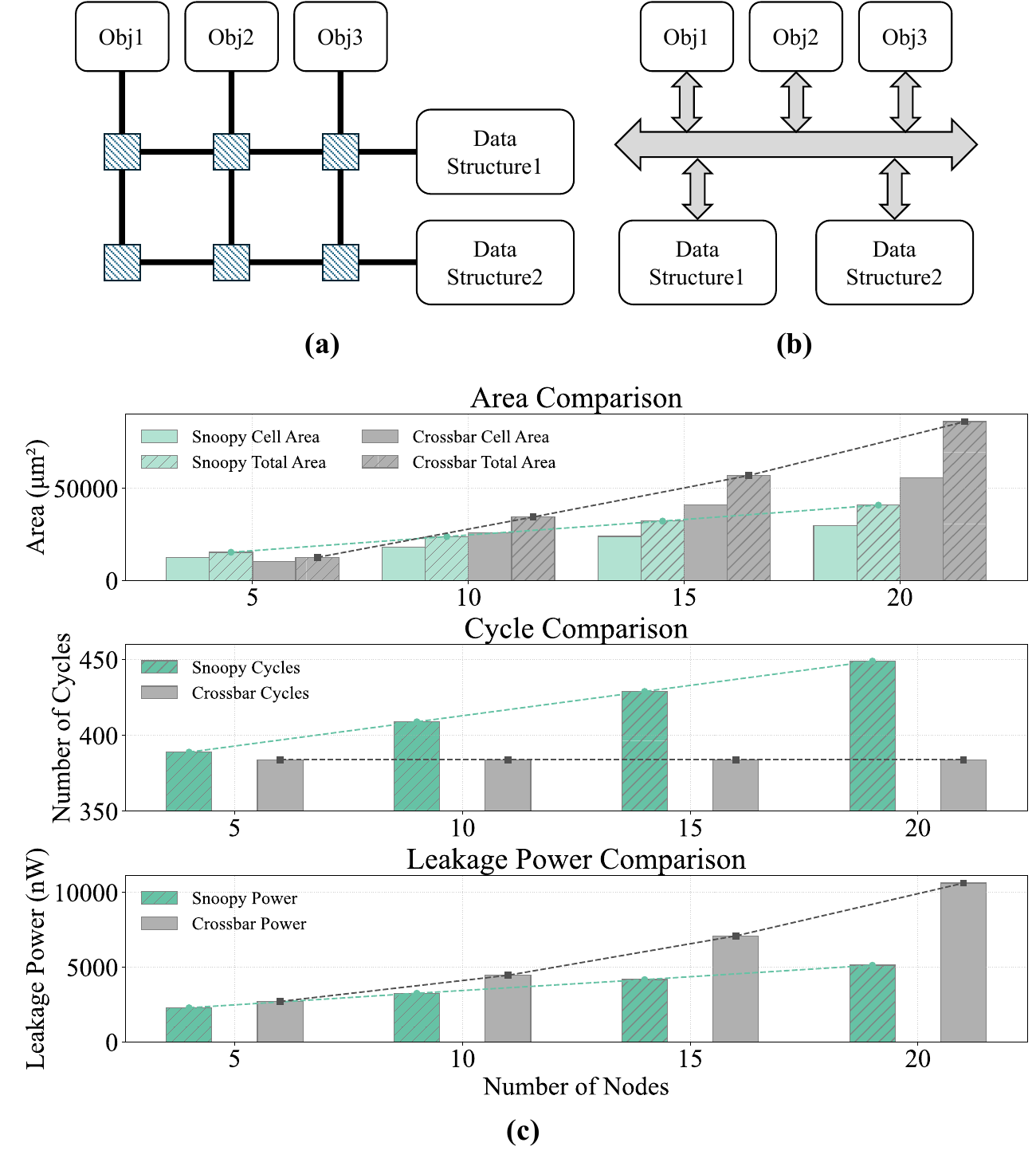}
  \caption{Comparison of snoopy bus and global crossbar architectures. 
(a) Global crossbar organization. 
(b) Snoopy bus organization. 
(c) Area, cycle count, and leakage power comparison between the two network architectures, evaluated using Intel 16nm technology.}

  \label{fig-snoopy-vs-crossbar}
  \vspace{-0.0in}

\end{figure}

\subsection{Network Architecture: Bridging Software Constructs and Physical Hardware}
\label{sec-network}
In this section, we examine how software-level object abstractions inform the design of the underlying hardware network. While establishing a single logical connection is straightforward, efficiently interconnecting multiple modular nodes is more challenging. Logical communication links, as defined by session-type sequences, cannot assume physical adjacency, since modules are distributed across vertical IPs. Therefore, once a communication candidate is selected by the framework, it must be instantiated via a network architecture that preserves the verified communication semantics while mapping logical specifications onto physically realizable hardware connections.

Our design adheres to three key principles. First, communication must remain pairwise, consistent with the software-level session abstractions. Second, the network must not perform additional computation, as this would alter the verified communication semantics and compromise equivalence. Lastly, the architecture must preserve behavioral correspondence between software-level objects and hardware-level implementations. By following these principles, we ensure that session-type communications verified at the software level are realized in the physical hardware.

We consider two representative network topologies, illustrated in Figure~\ref{fig-snoopy-vs-crossbar}: the snoopy bus and the crossbar network. The snoopy bus provides a natural mapping from session-type sequences to hardware. Since all edges reside on the same shared medium, a session sequence can be realized as a simple, sequential pairwise communication over shared wires~\cite{agarwal-snoopy-network-2009hpca, agarwal-snoopy-network-2009proceeding}. This approach preserves the order of session-type actions directly and does not allow concurrency. Its primary advantage is simplicity, low resource overhead, and the preservation of communication sequences similar to that of the original software system, which is an important property. However, the lack of concurrency can limit scalability as the number of communicating modules increases.
In contrast, a global crossbar relaxes ordering constraints by allowing multiple pairwise communications to proceed concurrently~\cite{pippenger-crossbar-network-2003tc, passas-crossbar-network-2012tc, sawchuk-crossbar-network-1987optical}. Each edge in the session-type graph can be mapped to a dedicated link, enabling greater concurrency and throughput at the cost of increased hardware resources. While more complex than the snoopy bus, the crossbar better supports modular and parallel interactions.

Using Intel 16nm technology, we implemented the required IP modules and network architecture to support a small CNN module. Figure\ref{fig-snoopy-vs-crossbar} illustrates how the relative efficiency of snoopy bus and crossbar networks depends on system size and complexity. For $N$ nodes, the snoopy bus scales linearly in area, $A_\text{snoopy} \sim O(N)$, while a full crossbar grows quadratically, $A_\text{crossbar} \sim O(N^2)$, due to dedicated links between all node pairs. In terms of communication latency, measured as the total cycles required to complete a CNN operation with identical inputs while varying the number of object nodes, the crossbar achieves near-constant cycles per transaction and benefits from concurrent transfers. In contrast, the snoopy bus experiences increasing latency with $N$ because all transactions share a single communication medium.
Regarding power, the snoopy bus exhibits lower static and dynamic consumption owing to reduced wire count and the absence of parallel paths. These trade-offs suggest that snoopy buses are preferable for area- and power-constrained designs with moderate concurrency, while crossbars are more suitable for latency-sensitive applications that require high parallelism.


\begin{figure}[t]

  \centering
  \includegraphics[width=1\linewidth]{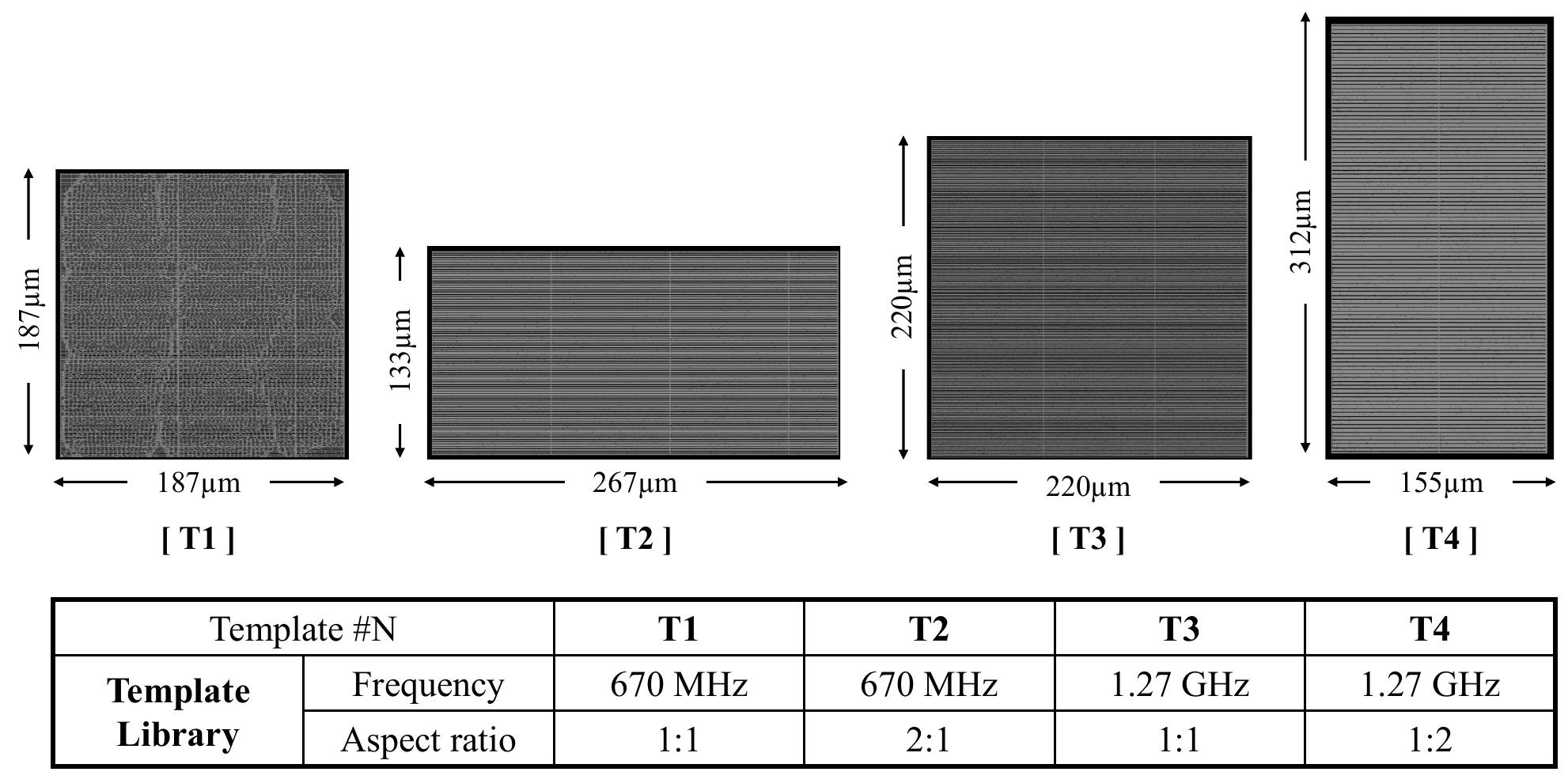}
  \caption{Template layout library of an average filter IP, generated under varying frequency and aspect ratio configurations in Intel 16nm technology. The library can be extended to include a range of architectural parameters (e.g., pipelining depth) and VLSI design parameters (e.g., multi-Vt libraries).}

  \label{fig-template-library}
  \vspace{-0.0in}

\end{figure}

\begin{figure}[t]

  \centering
  \includegraphics[width=0.95\linewidth]{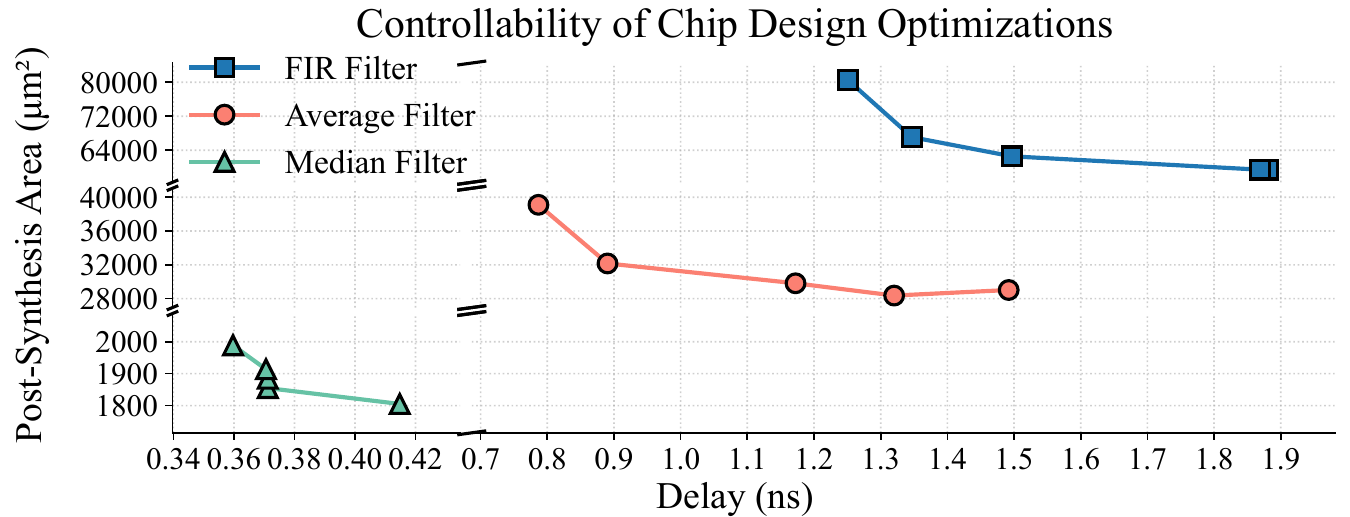}
  \caption{Characterization of three filter IPs with pipelining, offering design optimization space, synthesized in Intel 16nm technology.}

  \label{fig-design-space}
  \vspace{-0.0in}

\end{figure}


\begin{figure*}[t]

  \centering
  \includegraphics[width=0.88\linewidth]{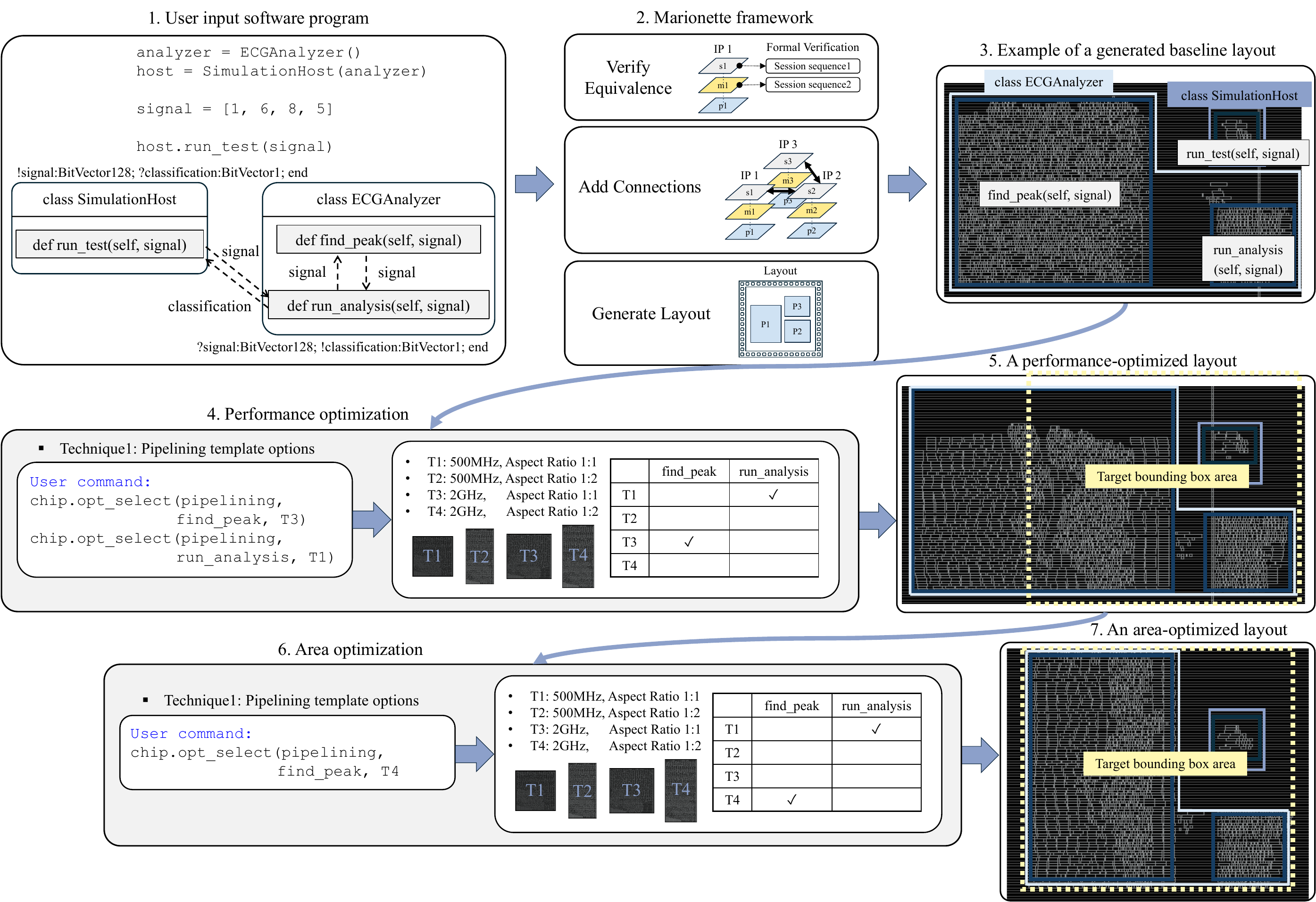}

  \caption{Iterative layout optimization guided by user commands and supported by the template library for design space exploration. Layouts are generated in Intel 16nm technology. The user passes their object-oriented program into Marionette and ses in the GUI the layout of the objects, and then uses a simple \texttt{opt\_select} interface to choose between physical layouts with different design metric characteristics for each IP block. Here, the user iterates within the GUI and attempts to fit all of their objects within the dotted yellow bounding box, thereby opening the black box of optimization for performance, area, and silicon utilization within software-first chip design.}

  \label{fig-optimization}
  \vspace{-0.0in}

\end{figure*}

\subsection{Design Optimization Under Constraints}
\label{sec-template-library}

Design optimization within Marionette is facilitated by a template library, as illustrated in Figure~\ref{fig-template-library} for a single vertical IP block. Each vertical IP block provides pre-verified physical layouts across multiple frequency and aspect ratio configurations, enabling users to select options that balance performance, area, and power according to their design goals.

\paragraph{Goals} As discussed in Section~\ref{sec-system-goals}, we seek to build working chips and a chip design optimization space for software-first users to explore, even at the cost of placing this space at a lower absolute range of design quality metrics. Producing working chips is very challenging, detail-oriented, and messy (e.g., alignment cells, physical isolation, di/dt decoupling caps). Our key insight is that significantly lowering the design quality target also drastically lowers the challenges of chip design (e.g., push-button flows, after initial expert technology-specific setup). We choose to make this tradeoff and then, at a low absolute range of metrics, create accessible, discrete, and templatized chip design optimization spaces.

\paragraph{Correctness by Construction}
A key property of Marionette is that designs are LVS- and DRC-clean by construction. Each template is independently verified to satisfy all layout design rules, including spacing, connectivity, and metal density. By hierarchically instantiating these templates, the framework ensures intra-block manufacturability, while inter-template connections follow predefined, rule-compliant routing tracks and grid structures, eliminating the need for ad-hoc routing corrections. Physical considerations such as signal integrity, decoupling capacitance, and cell placement are encapsulated within each pre-characterized template. As a result, composing templates does not introduce new violations, allowing users to optimize at the software level without manually managing low-level physical constraints.

\paragraph{Templatized Optimization}
Optimization in Marionette is discretized: users choose from pre-verified template floorplans and supported frequency options rather than performing continuous tuning. As illustrated in Figure~\ref{fig-design-space}, the template library enables controlled design optimization through techniques such as pipelining, register rebalancing, and clock gating. The optimization landscape in Figure~\ref{fig-design-space} guides the selection of appropriate template coverage. While this approach ensures correctness by construction, it demands careful management of template coverage because generating and verifying additional frequencies, aspect ratios, or specialized architectural variants remains time-consuming and resource-intensive.

\paragraph{Performance-Aware Architectural Features}
To enhance system throughput while maintaining physical correctness, templates incorporate architectural techniques such as pipelining and register rebalancing. Users can leverage the Marionette optimization flow to select and configure these strategies explicitly, trading off frequency, latency, and area.

\paragraph{Iterative Optimization Flow}
Figure~\ref{fig-optimization} illustrates Marionette’s iterative optimization process:
A user-provided software specification is first mapped to a set of module connections drawn from the template library.
Cross-layer equivalence verification ensures that these connections preserve the verified session-type semantics, producing a baseline layout.
Users may iteratively adjust template selections—for example, choosing higher-frequency configurations to improve performance or alternate aspect ratios to satisfy area constraints—gradually refining the layout.
Because each template is pre-verified and all interconnections follow the framework’s rules, both DRC and LVS correctness are preserved automatically throughout the optimization iterations. Moreover, functional correctness is maintained, as the selection and composition process inherently preserves the behavioral equivalence guaranteed by session-type verification.


\begin{table*}[t]
\centering
\small

\caption{Inference Latency (µs) Across Input Sizes and Microcontrollers}
\label{tbl-arduino}
\begin{tabular}{cccc|c|cccccc}

\toprule
\multicolumn{4}{c|}{\textbf{Input}} & \multicolumn{7}{c}{\textbf{Latency (µs)}} \\
\cmidrule(lr){1-4} \cmidrule(lr){5-11}
Testcase & Input Size & Module & MAC & Snoopy Bus & Uno & Nano BLE & Due & ESP32 & Portenta H7 & Teensy 4.1 \\
\midrule
TC1  &8x8  & Convolution & 324  & 0.5049 & 20.25 & 5.06 & 3.8571 & 1.35 & 0.675 & 0.54 \\
     &     & Relu \& Pool & 0   & 0.0126 & \~0 & \~0 & \~0 & \~0 & \~0 & \~0 \\
     &     & FC           & 9   & 0.0168 & 0.56 & 0.14 & 0.1071 & 0.037 & 0.0188 & 0.015 \\
     &     & Top          & 333 & \textbf{0.5371} & 20.81 & 5.2 & 3.9642 & 1.387 & 0.6938 & \textbf{0.555} \\
\midrule
TC2  &12x12 & Conv & 900  & 1.4000 & 56.25 & 14.06 & 10.7143 & 3.75 & 1.875 & 1.5 \\
     &       & Relu \& Pool & 0   & 0.0126 & \~0 & \~0 & \~0 & \~0 & \~0 & \~0 \\
     &       & FC   & 25  & 0.0406 & 1.56 & 0.39 & 0.2976 & 0.1042 & 0.0521 & 0.0417 \\
     &       & Top  & 925 & \textbf{1.4559} & 57.81 & 14.45 & 11.0119 & 3.8542 & 1.9271 & \textbf{1.5417} \\
\midrule
TC3  &14x14 & Conv & 1332 & 2.0154 & 81 & 20.25 & 15.4286 & 5.4 & 2.7 & 2.16 \\
     &       & Relu \& Pool & 0    & 0.0126 & \~0 & \~0 & \~0 & \~0 & \~0 & \~0 \\
     &       & FC   & 36   & 0.0559 & 2.25 & 0.56 & 0.4286 & 0.15 & 0.075 & 0.06 \\
     &       & Top  & 1456 & \textbf{2.0867} & 83.25 & 20.81 & 15.8572 & 5.55 & 2.775 & \textbf{2.22} \\
\midrule
TC4  &16x16 & Conv & 1764 & 2.7427 & 110.25 & 27.56 & 21 & 7.35 & 3.675 & 3.675 \\
     &       & Relu \& Pool & 0   & 0.0126 & \~0 & \~0 & \~0 & \~0 & \~0 & \~0 \\
     &       & FC   & 49  & 0.0741 & 3.06 & 0.77 & 0.5833 & 0.2042 & 0.5833 & 0.0817 \\
     &       & Top  & 1813 & \textbf{2.8322} & 113.31 & 28.33 & 21.5833 & 7.5542 & 4.2583 & \textbf{3.7567} \\
\bottomrule
\end{tabular}

\end{table*}

\section{Evaluation}
\label{sec-eval}

To our knowledge, our work is the first to investigate extending the object-oriented abstraction from software down to manufacturable chip layout. Given our audience of software-first users, we can compare the merits of our approach to the best hardware available to this audience, which primarily includes Arduino-style microcontrollers on cheaply accessible commercial boards. These are a typical first stop to integrating hardware into an innovative application for our target audience. We therefore evaluate Marionette chips in terms of their value as an Arduino extension (i.e., plug in your custom chip into an Arduino board).

\subsection{Latency Comparison between Arduino and Snoopy Bus-based CNN Modules}

To evaluate the performance of the Marionette framework for its primary target users --- software engineers requiring chip prototyping and students learning chip design --- we compared the latency of CNN modules implemented on Arduino boards and a Snoopy bus-based hardware network. 

For the CNN computation, four IP modules were connected: convolution (Conv), ReLU, pooling (Pool), and fully connected (FC). Input sizes were progressively increased to raise the total number of MAC operations, allowing latency comparisons under varying computational loads. The baseline setup used for experiments is: input size $8\times8$, kernel size $3\times3$, Conv output $6\times6$, ReLU $6\times6$, Pool $3\times3$, and FC with 9 MACs.

The Arduino platforms evaluated included Uno, Nano BLE, Due, ESP32, Portenta H7, and Teensy 4.1. For this comparison, we assumed an idealized scenario in which each board could perform one MAC operation per clock cycle, ignoring the latency contributions from ReLU and Pool operations. This simplification focuses the comparison on DSP-intensive workloads. In practice, each MAC operation may take multiple cycles, and non-MAC operations (e.g., memory accesses for ReLU and Pool) would introduce additional latency.

The Snoopy bus network was implemented using Intel 16,nm technology. Despite its bus-based topology, which could introduce serialization delays, it achieved lower latency than even the fastest Arduino board (Teensy 4.1) under equivalent computational loads, as summarized in Table~\ref{tbl-arduino}. Table~\ref{tbl-arduino} reports measured inference latencies for varying input sizes across microcontroller platforms and the Snoopy bus-based CNN implementation. As the input size and total MAC operations increase (TC1 to TC4), all platforms show higher latency; however, the Snoopy bus scales more efficiently. For instance, Teensy 4.1 latency increases from 0.555,$\mu$s (TC1) to 3.7567,$\mu$s (TC4), whereas the Snoopy bus grows more modestly from 0.5371,$\mu$s to 2.8322,$\mu$s.
Slower boards, such as Uno and Nano BLE, incur significantly higher latencies (e.g., Uno reaches 113.31,$\mu$s in TC4), underscoring the performance benefits of a dedicated hardware network for CNN modules. The Snoopy bus efficiently connects the four CNN IP modules (Conv, ReLU, Pool, FC) while minimizing communication overhead, yielding superior performance for MAC-dominated workloads. In this evaluation, ReLU and Pool operations were excluded due to negligible MAC contribution. While memory access and control logic introduce additional latency in real implementations, the relative performance trend compared to Arduino boards remains consistent, as these overheads are minor relative to total MAC processing time.
Overall, the Snoopy bus demonstrates sublinear latency growth with increasing MAC count, in contrast to the linear or superlinear scaling observed in Arduino boards due to limited parallelism and clock constraints. These results confirm the Marionette framework's effectiveness for rapid prototyping of CNN modules, providing predictable, low-latency execution. The Snoopy bus network is sufficient to provide a compelling alternative to conventional Arduino-based prototyping, particularly for MAC-intensive applications.

Ultimately, we provide a path for software-first chip designers to describe an object-oriented system, visualize a hardware layout that has been formally guaranteed to be equivalent by Marionette and retains the object-oriented abstraction for mental continuity, manipulate aspects of chip design via replacing each object's physical template (performance, area, aspect ratio), and finally producing a manufacturable chip that outperforms readily available Arduino-style hardware.


\section{Related Work}
\label{sec-related}
Recent efforts have aimed to lower the barriers to chip design~\cite{anand-skip-asplos2024, olofsson-zeroasic-2022, guthaus-nsf-workforce-dev-arxiv2023, venn-tinytapeout-2024, zero-asic-web, openroad-web, han-shakeflow-interface-combinators}. Community-driven initiatives like Tiny Tapeout~\cite{venn-tinytapeout-2024} and Zero ASIC~\cite{zero-asic-web} provide accessible entry points for students and non-experts, supporting small-scale and modular design projects. From the toolchain perspective, OpenROAD~\cite{openroad-web} offers a fully automated open-source digital flow, aiming for RTL-to-GDSII tapeout within 24 hours~\cite{ajayi-openroad-dac2019}. More recently, AI- and LLM-driven approaches have emerged~\cite{stelios-ai-chip-tutorial-hotchip2024, firouzi-chipmind-2025vts, blocklove-chipchat-arxiv2023}. ChipMind~\cite{firouzi-chipmind-2025vts} introduces a modular agent-based framework automating both digital and analog design, improving prototyping efficiency and enabling power, performance, area optimization with minimal human intervention. Chip-Chat~\cite{blocklove-chipchat-arxiv2023} demonstrated the first chip fabricated entirely from HDL generated via natural-language interaction with GPT-4, highlighting the potential for democratizing chip design. In contrast, our approach directly compiles software programs into hardware, preserving program semantics without relying on intermediate natural-language abstractions.  

These efforts share the common goal of making chip design more accessible through software-like abstractions or natural language~\cite{collini-c2hlsc-2025, pilato-bridging-gap-hw-sw-2018apc, king-sw-driven-hw-2015acm}. Traditional high-level synthesis (HLS) allows software algorithms to be translated into hardware~\cite{ye-hls-2023, liu-hls-reuse-date2012, hoe-gaa-synth-iccad2003, cong-hls-fpga-2022, choi-hls-flash-2020}, and languages like Chisel~\cite{bachrach-chisel-dac2012, schoeberl-chisel-book-2019} enable higher-level hardware construction. However, HLS and similar tools do not fully preserve software semantics in hardware and rarely guide layout-level optimizations. Our approach explicitly addresses these limitations, ensuring semantic preservation while supporting software-level layout-aware design. 

Research on modular chip networks has also advanced~\cite{passas-xbar-nocs2010, lee-noc-prototype-tvlsi2006, kumar-concentration-nocs2009, lee-noc-variable-freq-2009elsevier,beausoleil-noc-ieee2008,chang-noc-ieee2001,kim-noc-micro2007,Zheng-chiplet-noc-DAC2020,Mishra-heter-noc-DAC2013, duraisamy-noc-vlsi2017}. Zheng et al.~\cite{Zheng-chiplet-noc-DAC2020} propose flexible chiplet-based systems for efficient inter-chip communication and heterogeneous core integration, while Mishra et al.~\cite{Mishra-heter-noc-DAC2013} optimize NoC topologies for performance and energy efficiency. Unlike these high-performance-oriented designs, our work targets software-driven architectures: we implement a snoopy bus network aligned with session-type sequences, enabling predictable and verifiable communication patterns, and compare it against conventional microcontroller-based approaches.


\section{Conclusion}
\label{sec-conclusion}

We present a software-first chip design methodology that maps object-oriented software communication into hardware and manufacturable layouts while preserving software abstractions and ensuring protocol correctness via session types. The framework offers a formal software-hardware correspondence, an equivalence checker without false positives, and modular automated layout generation supporting design tradeoffs. Prototype evaluation shows a snoopy bus–based Marionette network as a viable alternative to Arduino prototyping. Despite current limitations, this approach lowers the barrier for software developers, with future work aimed at refining verification, optimization, and extending application support.


\bibliographystyle{IEEEtranS}
\bibliography{refs}

\end{document}